\documentclass[pdftex,twocolumn,epjc3]{svjour3}

\smartqed  

\RequirePackage{graphicx}
\RequirePackage{mathptmx}      
\RequirePackage{flushend}
\RequirePackage[numbers,sort&compress]{natbib}
\RequirePackage[colorlinks,citecolor=blue,urlcolor=blue,linkcolor=blue]{hyperref}

\journalname{Eur. Phys. J. C}

\usepackage{amsmath,amssymb}
\usepackage{color}
\usepackage{slashed}

\def\hyph{-\penalty0\hskip0pt\relax}

\newcommand{\skipthis}[1]{}
 \def\Eq#1{Eq.~(\ref{#1})}
\def\Fig#1{Fig.~\ref{#1}}

\begin{document}
\title{Real-Time Correlation Functions in the $O(N)$ Model\\from the Functional Renormalization Group}
\author{Kazuhiko Kamikado\thanksref{e1,addr2}
        \and
        Nils Strodthoff\thanksref{e2,addr3,addr4} 
        \and
        Lorenz von Smekal\thanksref{e3,addr3,addr5}
        \and
        Jochen Wambach\thanksref{e4,addr3,addr6}
}        
\thankstext{e1}{e-mail: kazuhiko.kamikado@riken.jp,}
\thankstext{e2}{e-mail: n.strodthoff@thphys.uni-heidelberg.de}
\thankstext{e3}{e-mail: lorenz.smekal@physik.tu-darmstadt.de}
\thankstext{e4}{e-mail: jochen.wambach@physik.tu-darmstadt.de}

\institute{
		  Theoretical Research Division, Nishina Center, RIKEN, Saitama 351-0198, Japan\label{addr2}
          \and
          Institut f\"{u}r Kernphysik, Technische Universit\"{a}t Darmstadt, 64289 Darmstadt, Germany\label{addr3}
          \and
          Institut f\"{u}r Theoretische Physik, Universit\"{a}t Heidelberg, 69120 Heidelberg, Germany\label{addr4}
          \and
          Institut f\"{u}r Theoretische Physik, Justus-Liebig-Universit\"{a}t, 35392 Giessen, Germany\label{addr5}
          \and
          GSI Helmholtzzentrum f\"{u}r Schwerionenforschung GmbH, 64291 Darmstadt, Germany\label{addr6}
}

\date{Received: date / Accepted: date}

\maketitle
\begin{abstract}
 In the framework of the functional renormalization group (FRG) we
 present a simple truncation scheme for the computation of real-time 
 mesonic $n$-point functions, consistent with the derivative expansion 
 of the effective action. Via analytic continuation on the level of the
 flow equations we perform calculations of mesonic spectral
 functions in the scalar $O(N)$ model, which we use as an exploratory
 example. By investigating the renormalization-scale dependence of the 2-point functions
 we shed light on the nature of the sigma meson, whose spectral properties
are predominantly of dynamical origin. 
 \end{abstract}

 \section{Introduction}
 Real-time observables in strongly-interacting systems
 often pose major challenges for theoretical calculations. Spectral  
 functions are key to many such observables. They provide information
 on quasi-particle spectra and collective excitations of the
 system. The spectral functions 
 of the energy-momentum tensor moreover allow to extract transport
 coefficients via the Kubo formulas
 \cite{Aarts:2002cc,Meyer:2008gt,Taylor:2010,Meyer:2011gj} and thus
 to study macroscopic properties of strongly-interacting matter.

 In this letter we study spectral functions of the elementary sigma and
 pion correlations in the $O(N)$ model with the functional
 renormalization group (FRG). FRG methods
 \cite{Wegner:1972ih,Polchinski:1983gv,Wetterich:1992yh} have been well
 developed over the last thirty years. Although various formulations
 exist, their underlying physical principles are always the same. The
 equations of motion are described as functional differential equations
 of a scale-dependent generating functional and functional derivatives
 of it.
 
 Here we employ the approach pioneered by Wetterich
 \cite{Wetterich:1992yh} with the scale-dependent effective average
 action as the central object, which generalizes the usual effective
 action (see
 \cite{Berges:2000ew,Polonyi:2001se,Pawlowski:2005xe,Schaefer:2006sr,Gies:2006wv}
 for a general introduction). By integrating the Wetterich equation
 which describes the evolution of the effective average action with the
 RG scale $k$, one obtains the full quantum effective action in the
 limit $k\to 0$. The evolution equations for the $n-$point functions can
 be obtained from that of the effective action via functional
 derivatives.

 A typical problem which one encounters in functional ap\-proaches is the fact
 that the flow equations for $n$-point functions involve information
 on up to $(n+2)$-point functions thereby leading to an infinite
 tower of coupled evolution equations. Practical applications thus
 require truncations in order to obtain a closed system of equations
 to solve. One frequently used truncation scheme is the derivative
 expansion based on expanding the effective action in terms of
 derivatives. The leading order in this expansion is the so-called
 local potential approximation (LPA). Despite their simplicity, derivative 
 expansions in general and the LPA in particular, have been applied
 successfully to a broad range of physical systems and critical phenomena. 
 
 In this letter we present an extension of the derivative
 expansion to obtain real-time 2-point correlation functions consistent
 with the underlying truncation for the effective potential. Starting
 from the flow equation for 2-point functions we employ LPA vertices
 to obtained a closed system of flow equations 
 involving as the only input the scale\hyph dependent effective potential. A
 consistent truncation scheme for the calculation of the Euclidean
 2-point correlators at finite external momentum is, however, only the first
 step towards a calculation of real-time quantities such as spectral
 functions or transport coefficients. 

 The crucial second step, a common challenge in Euclidean approaches to
 thermal quantum field theory, is the analytic continuation of
 the external momentum to Minkowski space-time. At present, real-time
 correlation functions are usually  
 either reconstructed from their Euclidean analogues using Pad\'{e}
 approximants or by maximum entropy methods.  Alternatively, they can be 
 calculated directly in Minkowski space-time by an analytical
 continuation at the level of the flow equations
 \cite{Strodthoff:2011tz,Kamikado:2012bt}. In this respect, our approach
 is similar to that of Ref.~\cite{Floerchinger:2011sc} in which 
 more refined truncations in real time were proposed to include
 effects of the back-coupling of a non-trivial propagator on the
 effective potential and the wavefunction renormalization. Actual
 solutions, however, require an Ansatz for the form of the propagator
 and its singularity structure. Our approach, on the other hand, does
 not rely on assuming a certain spectral shape,  
 but it ignores back-coupling effects at present.
 
 In the following, we concentrate on the calculation of spectral
 functions within the $O(N)$ model. The analysis of the $O(N)$ scalar
 model, which is frequently used as a chiral effective model for QCD
 $(N=4)$, has a long history. Spectral functions in $O(N)$ models have
 been calculated for example in optimized perturbation theory
 \cite{Hidaka:2002xv,Chiku:2000eu,Chiku:1998kd,Chiku:1997va} 
 by taking into account the $\sigma \rightarrow \pi + \pi$
 process. The critical exponents have been evaluated in
 \cite{Chiku:2000eu}. Although quantum  
 fluctuations are included via a resummed loop expansion, the critical exponents
 remain at their mean-field values. To overcome this limitation a
 more suitable resummation scheme is required. Such a framework is
 provided by the FRG, where the application of the derivative expansion
 leads to very accurate results, \textit{e.g.}, for critical exponents in $O(N)$
 models in higher orders of the derivative expansion
 \cite{Berges:2000ew,Canet:2003qd,Litim:2010tt} but to a surprising
 degree already also in the LPA \cite{Bohr:2000gp,Litim:2002cf}.

 Spectral functions have been calculated with the FRG in
 non-relativistic models using different truncation schemes including
 the ``BMW'' approximation \cite{Blaizot:2005wd, Blaizot:2006vr}, 
 vertex expansions or derivative expansions
 \cite{Dupuis:2009,Sinner:2009,Schmidt:2011}. In all these cases,
 however, the spectral functions were reconstructed from Euclidean
 2-point correlators via analytic continuation using Pad\'{e}
 approximants. Here we employ a simple LPA for the effective action
 which we then use to solve analytically continued flow equations 
 for the 2-point functions at real frequency. For simplicity, we
 restrict ourselves to the zero\hyph temperature case in this work. In
 principle it is possible to extend the proposed approximation scheme
 to finite temperature and finite chemical potential or to include fermions 
 \cite{Strodthoff:2011tz,Kamikado:2012bt}.
 
 \section{Formulation}
  \subsection{Functional RG}
  The functional renormalization group provides a powerful
  non-perturbative tool, especially for the study of critical
  phenomena. It describes the evolution of a
  scale-dependent effective average action $\Gamma_k$ from the
  microscopic bare action, specified at an ultra-violet (UV) scale
  $k=\Lambda$, to the corresponding full quantum effective action for
  $k\to 0$ in terms of a functional differential equation 
  \cite{Wetterich:1992yh},
  \begin{equation}
   \partial_k \Gamma_k  = \frac{1}{2}{\rm Tr}\!\left[\partial_k
						R_{k}(\Gamma^{(2)}_k+R_{k})^{-1}\right],
   \label{eq:floweq}
  \end{equation}
  where $\Gamma_k^{(n)}$ denotes the $n^\text{th}$ functional derivative
  of $\Gamma_k$ with respect to the fields, and $R_k$ is 
  a suitable regulator function. The
  trace is taken over the internal degrees of freedom and momentum space.
  \begin{figure}[ht]
   \centering
   \vspace*{.2cm}
   \includegraphics[width=0.45\columnwidth]{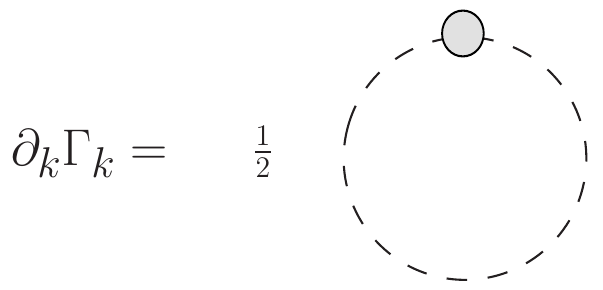}
   \vspace*{.2cm}
   \caption{Diagrammatic representation of the flow equation for the
   effective action of the $O(N)$ model. Dashed lines represent 
   scale-dependent boson propagators and gray filled circles
   correspond to regulator insertions $\partial_k R_k$.}  \label{fig:picture_flow}
  \end{figure}
  \Fig{fig:picture_flow} shows a diagrammatic representation of the flow
  equation (\ref{eq:floweq}). Note that although the flow equation has a
  one-loop structure, it is an exact equation, involving full
  (scale-- and field-dependent) propagators.  

  The precise form of the regulator function $R_{k}$ is not fixed but
  limited only by certain general requirements \cite{Berges:2000ew}. 
  A convenient choice for our purposes is the sharp regulator
  \begin{equation}
   R_{k} = (k^2 - \vec{q}^2) \theta (k^2 - \vec{q}^2),
    \label{eq:regulators}
  \end{equation}
  which is the three-momentum analogue of the LPA\hyph optimized regulator
  \cite{Litim:2001up}. The introduction of the regulator suppresses the
  propagation of field modes with momenta smaller than the renormalization scale
  $k$. Integrating the flow equation (\ref{eq:floweq}) from the
  bare classical action at $k=\Lambda$ down to $k=0$ yields
  the full quantum effective action which includes quantum fluctuations
  from all momentum modes. Taking $n$ functional derivatives of the flow
  equation (\ref{eq:floweq}) one obtains the flow equations for the
  $n$-point correlation functions which, in turn, contain up to
  $(n+2)$-point correlation functions.
  \subsection{Flow equation for the effective action}
  The derivative expansion corresponds to an expansion of the effective
  action in terms of gradients
  \begin{equation}
    \begin{split} 
   \Gamma_{k}[\phi] &=  \int \!dx \; \Big\{  U_k(\phi^2)-c \sigma \\
    & \hskip 1cm 
   +\frac{1}{2}Z_{1,k}(\phi^2) (\partial \phi)^2 + Z_{2,k}(\phi^2) \;
    (\partial \phi)^4 + \cdots \Big\} \, ,
  \end{split}
  \label{eq:ansatzeffectiveaction}
  \end{equation}
  where the effective potential $U_k$ and the wavefuction
  renormalization factors $Z_{i,k}$ are parametrized in terms of the
  $O(N)$ invariant $\phi^2 = \phi_i \phi^i$ and an explicit symmetry
  breaking is induced by the $c\sigma$ term giving rise to a finite pion
  mass. To study the $O(N)$ symmetry breaking one decomposes $\Gamma^{(2)}_k$
  for constant background $\phi_i$ into its transverse ($\Gamma_{\pi}$)
  and longitudinal ($\Gamma_{\sigma}$) components in field space
  \begin{equation}
   \label{eq:decomposition}
    \begin{split}
     \Gamma_{k,ij}^{(2)}(\phi^2,p) = \Gamma^{(2)}_{k,\pi}(\phi^2,p)\left(\delta_{ij} - \frac{\phi_i
     \phi_j}{ \phi^2}\right) + \Gamma^{(2)}_{k,\sigma}(\phi^2,p) \frac{\phi_i \phi_j}{ \phi^2}.
    \end{split}
  \end{equation}
  Inserting the Ansatz~(\ref{eq:ansatzeffectiveaction}) into the flow
  equation (\ref{eq:floweq}) and evaluating it for a constant field
  configuration yields the corresponding flow equation for the effective
  potential. In the simplest case of the LPA which corresponds to
  setting $Z_{1,k} = 1$ and $Z_{i,k} = 0$ for $i>1$ one is left with 
  solving the flow equation for the effective potential
  which reads
  \begin{equation}
   \begin{split}
    \frac{\partial U_k}{\partial k}& = \frac{1}{2} I^{(1)}_\sigma +
    (N-1) \frac{1}{2}I^{(1)}_\pi.
   \end{split}
   \label{eq:floweqeffectivepotential}  
  \end{equation} 
  Here we have defined the loop functions $I^{(n)}_{i}$ as
  \begin{equation}
   I^{(n)}_{i}\equiv {\rm Tr}_q\left[ \partial_k R_k(q)
			    \left(\frac{1}{\Gamma_{k,i}^{(2)}+R_k}\right)^n_{q}  \right],
  \end{equation}
  where the trace runs over the loop-momentum $q$. For the regulator choice in
  \Eq{eq:regulators}, $I_i^{(1)}$ and $I_i^{(2)}$ are given by
  \begin{equation}
   I^{(1)}_{{i}} = \frac{k^4}{3 \pi^2} \frac{1}{2
   E_i}\quad\text{and}\quad I^{(2)}_{k,i} =\frac{k^4}{3 \pi^2}
   \frac{1}{4E_i^3}\;,
  \end{equation}
  where $E_\pi=\sqrt{k^2+2 U'}$, $E_\sigma=\sqrt{k^2+2 U'+4 U''\phi^2}$
  and $U'= \frac{\partial U}{\partial (\phi^2)}$. The curvatures $2U'$
  and $2U' + 4U''$ at the minimum of the effective potential 
  yield the screening masses squared of pion
  ($m^\mathrm{scr}_{\sigma}{}^2$) and sigma meson ($m^\mathrm{scr}_{\pi}{}^2$),
  respectively. They are defined as the spacelike limits $p\to 0$ of the
  pion and sigma 2-point functions, \textit{i.e.}, the inverse of the
  corresponding susceptibilities.      
  
  \subsection{Flow equation for the 2-point functions}
\begin{figure}[ht]
   \centering
   \vspace*{.2cm}
   \includegraphics[width=0.9\columnwidth]{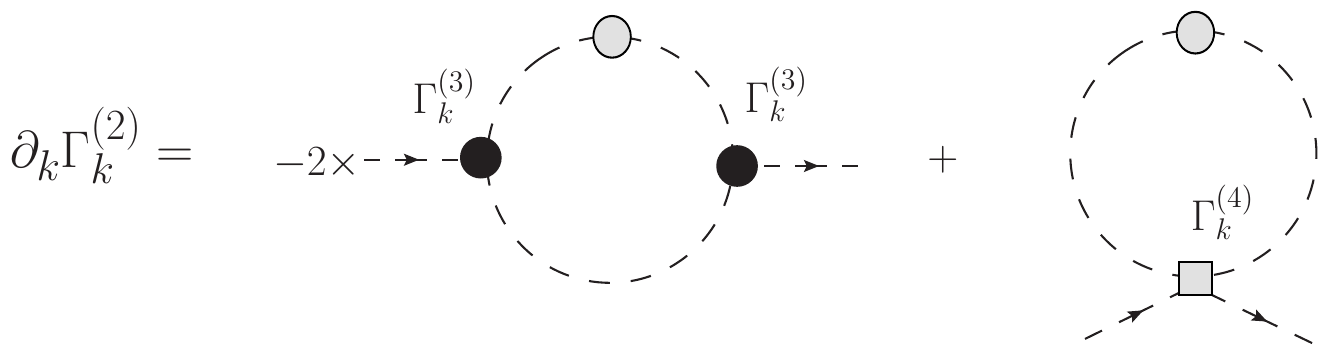}
   \vspace*{.2cm}
   \caption{Diagrammatic representation of the flow equation for a
   mesonic 2-point function in the $O(N)$ model. Dashed lines represent 
   scale-dependent boson propagators and gray filled circles
   correspond to regulator insertions $\partial_k R_k$. Black filled
   circles and gray filled squares denote the 
   scale-dependent 3- and 4- point vertex functions.}  \label{fig:picture_2pt}
  \end{figure}

  Applying two functional derivatives to the flow equation in
  \Eq{eq:floweq}, we obtain the exact flow equations for 2-point
  correlation functions. Their diagrammatic forms for the pion and the
  sigma in the $O(N)$ model are shown in \Fig{fig:picture_2pt}. These
  flow equations contain scale-dependent 3- and 4-point functions.
  In order to close the infinite tower of functional equations for the $n$-point
  functions, truncations are required. A possible scheme has been 
  proposed in Refs.~\cite{Blaizot:2005wd, Blaizot:2006vr} (BMW). The
  idea is that, 
  because of the insertion of the cutoff function, the external momentum
  dependence of the scale-dependent 3- and 4-point functions has a
  weaker effect than their dependence on the momenta of internal lines
  containing the loop-momentum. One thus expands the 3- and 4-point
  functions in their external momenta in this scheme.

  Here we choose a simpler truncation which is a natural extension of
  the derivative expansion. For this we replace the 3- and
  4-point functions by their corresponding scale dependent but, 
  at the leading order derivative expansion, momentum independent
  forms as obtained from the flow equation  
  for the effective potential, \textit{i.e.},
  \begin{equation}
   \begin{split}
    \Gamma^{(3)}_{ijm} &\rightarrow \frac{\partial^3
    \Gamma_{k}}{\partial
    \phi^m \partial \phi^j \partial \phi^i} \\
    \Gamma^{(4)}_{ijmn} &\rightarrow \frac{\partial^4 \Gamma_{k}}{\partial
    \phi^n \partial \phi^m \partial \phi^j \partial \phi^i} .
    \label{eq:ansatz3and4ptfn}
   \end{split}
  \end{equation}
  We then obtain the flow equations for the
  pion/sigma 2-point functions as follows:
  \begin{equation}
   \label{eq:flow2ptpisi}
    \begin{split}
     \partial_k \Gamma_{k,\pi}^{(2)}& = (J_{\sigma \pi}+J_{\pi \sigma})(4U'' \phi)^2
     \\
     &-\frac{1}{2} I^{(2)}_{\pi}4U'' (N+1) -\frac{1}{2} I^{(2)}_{\sigma}(4U'' + 8U'''\phi^2) \\
     \partial_k \Gamma_{k,\sigma}^{(2)}& = J_{\sigma \sigma}
     (12 U''\phi+8U'''\phi^3)^2 + J_{\pi \pi}(4U''\phi)^2(N-1) \\
     &-\frac{1}{2} I^{(2)}_{\sigma} (12U''+48U'''\phi^2 +16U''''\phi^4) \\
     &-\frac{1}{2} I^{(2)}_{\pi}(N-1) (4U''+8U'''\phi^2),
    \end{split}
  \end{equation}
  with loop functions $J_{ij}$ defined as
  \begin{equation}
   J(p)_{ij}\equiv {\rm Tr}_q \left[\partial_k R_k(q)
			   \left(\frac{1}{\Gamma_{k,i}^{(2)}+R_k}\right)_{p+q}\left(
			    \frac{1}{\Gamma_{k,j}^{(2)}+R_k}  \right)^2_{q}\right],
  \end{equation}
  where the trace runs over the momentum $q$. Note that only the symmetric
  components of $J$ are needed in \Eq{eq:flow2ptpisi}. In LPA, for vanishing
  spatial external momentum $\vec{p}$, and with the regulator
  in Eq.~(\ref{eq:regulators}), these are obtained as 
  \begin{equation}		
   \begin{split}
    &J(p_0)_{ij} + J(p_0)_{ji} = \\
    &\;\;\;\;\frac{k^4}{3\pi^2}  \frac{(E_i+E_j)^3(E_i^2+E_iE_j+E_j^2)
    +(E_i^3+E_j^3) p_0^2}{4
    E_i^3E_j^3\left(p_0^2+(E_i+E_j)^2\right)^2}\;.
    \label{eq:expression_J}
   \end{split}
  \end{equation}

  \begin{figure}[t]
   \centering \includegraphics[width=0.97\columnwidth]{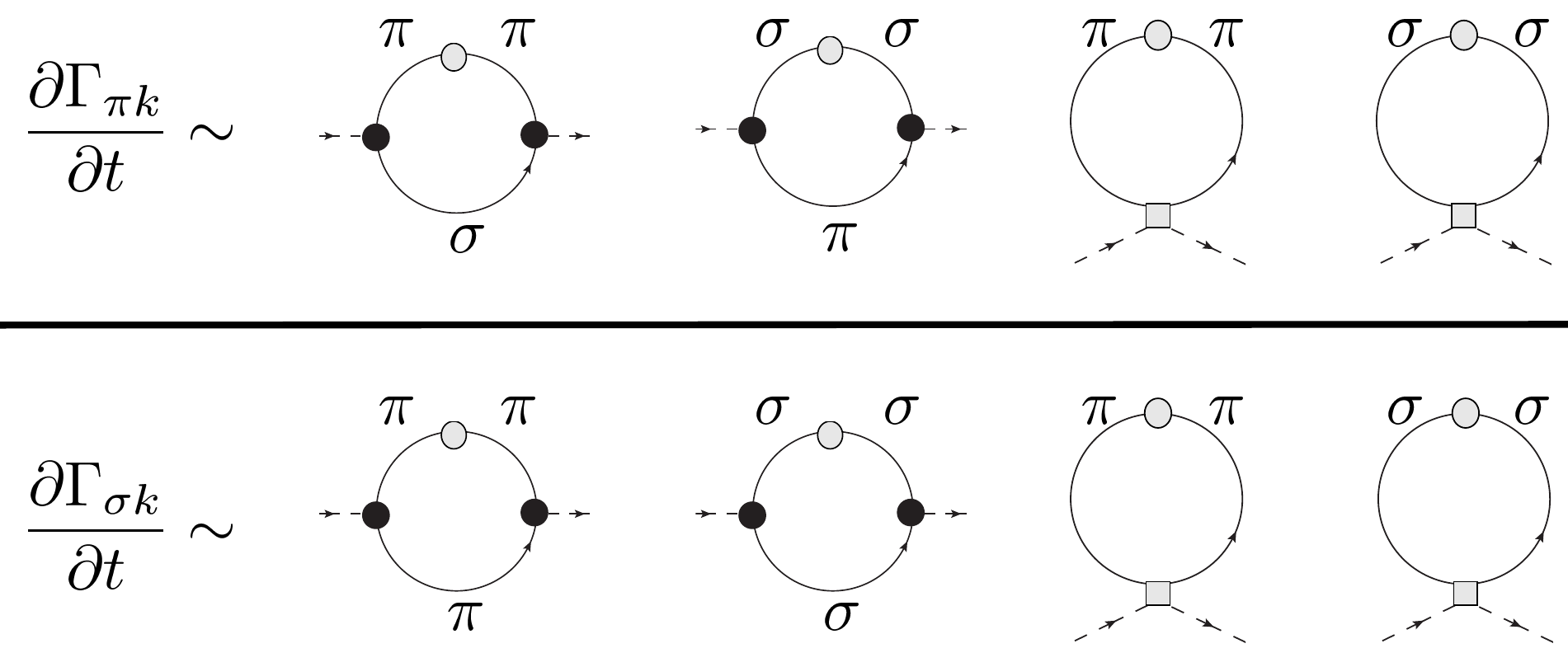}
   \caption{Diagrammatic representation of the contributions to the flow
   of the 2-point functions.  Full gray circles correspond to insertions of
   $\partial_k R_k$. Black circles and gray squares indicate the
   scale-dependent 3- and 4- point vertex functions,  here obtained from
   (\ref{eq:ansatz3and4ptfn}).}  \label{fig:contributions}
  \end{figure}

\noindent
  In \Fig{fig:contributions}, we show diagrammatically the contributions
  to the pion/sigma 2-point functions. Although they look like
  1-loop contributions, the internal lines represent full
  scale-dependent propagators which we evaluate for the constant field
  value at the minimum of the effective potential. These propagators 
  correspond to resummations of perturbative diagrams. For the pion 2-point
  function the process $\pi 
  \rightarrow \pi + \sigma$ is included and for the sigma 2-point function
  the processes $\sigma \rightarrow \sigma + \sigma$ and $\sigma
  \rightarrow \pi + \pi$. Other contributions, such as $\sigma
  \rightarrow \sigma + \pi$, would break
  the residual $O(N-1)$ symmetry of the broken phase and are excluded.

  In the following we discuss the solutions to
  Eqs.~(\ref{eq:flow2ptpisi}) at zero temperature. 
  Note that then, no term proportional to
  $1/[p_0^2+(E_{\sigma}-E_{\pi})^2]$ arises because Landau
  damping, \textit{i.e.}, the absorption of an on-shell particle from
  the heat bath, does not occur.  

  We reemphasize that the correlation functions at vanishing external
  momentum in our truncation are consistent with
  the effective action since the flow equations satisfy
  \begin{align}
   \partial_k\Gamma^{(2)}_{k,\pi}(p=0)&=2\partial_k
   U_k'\,,\label{eq:2pf_initialpion}\\
   \partial_k\Gamma^{(2)}_{k,\sigma}(p=0)&=2
   \partial_k U_k'+4 \partial_k U_k'' \phi^2\,.
   \label{eq:2pf_initialsigma}
  \end{align}
  This means that the equivalence between the static screening masses
  from the spacelike $p\to 0$ limit of the 2-point functions
  and from the curvatures of the effective potential at its minimum is
  manifest in our truncation. Moreover, note
  that Eq.~(\ref{eq:2pf_initialpion}) automatically ensures the
  existence of a dynamical Nambu-Goldstone boson in the chiral limit. Thus our
  truncation scheme is also ``symmetry conserving.''
  
  We solve the flow equation for the real and imaginary parts of the retarded
  2-point correlators. These are obtained from \Eq{eq:flow2ptpisi} via the
  analytic continuation
  \begin{equation}
   \Gamma^{(2) R}_{k,j} (\omega) \equiv \Gamma^{(2) R}_{k,j}
    (p_0=-i(\omega+i \epsilon),\vec p =0)\, , \quad \text{for}\;\;
    j=\pi,\sigma
    \label{eq:analytic_continuation}
  \end{equation}
  which is taken before the evaluation of the flow equations.

  The spectral functions are proportional to the imaginary parts of the
  retarded propagators or more explicitly, for $\omega>0$, 
  \begin{equation}
   \label{eq:spectralfn} \rho_i(\omega) = - \tfrac{1}{\pi} \tfrac{{\rm
    Im } \, \Gamma^{(2) R}_{i}(\omega)}{\left({\rm Re}\, \Gamma^{(2)
				 R}_{i}(\omega)\right)^2 +\left({\rm Im}\, \Gamma^{(2) R}_{i}(\omega)\right)^2
				 }\, , \quad i=\pi,\sigma,
  \end{equation}
  which are understood to be evaluated in the limit $\epsilon \to
  0^+$. In the numerical evaluations we have to keep a small but
  finite positive imaginary part in the external energy, however.

 \section{Numerical results}
  \subsection{Numerical method}
  In order to solve the flow equations numerically, we employ the
  Taylor-expansion method.  We expand the flow equations
  (\ref{eq:floweqeffectivepotential}) and (\ref{eq:flow2ptpisi}) by
  expanding $U_k$ and the real or imaginary part of $\Gamma^{R}_{k,i}$
  ($i=\sigma,\pi$) around the scale dependent minimum:
  \begin{equation}
   \begin{split}
    U_k &= \sum_{n=0}^{K} a_{n,k} (\phi^2 -
    \phi_{k}^2)^n \, ,\\
    {\rm Re}\,\Gamma^{(2) R}_{k,i}(\omega)& = \sum_{n=0}^{L}
    b^i_{n,k}(\omega) (\phi^2 - \phi_{k}^2)^n \, , \\
    {\rm Im}\,\Gamma^{(2) R}_{k,i}(\omega)& = \sum_{n=0}^{L}
    c^i_{n,k}(\omega) (\phi^2 - \phi_{k}^2)^n \, .
   \end{split}
  \end{equation}
  The flow equations for the effective potential and the 2-point
  functions then translate into flow equations for the expansion
  coefficients $a_{n,k}$, $\phi_{k}^2$, $b^i_{n,k}(\omega)$ and
  $c^i_{n,k}(\omega)$ in the usual way. Choosing $L=K-1$ ensures that
  the consistency condition (\ref{eq:2pf_initialpion}) between the pion
  2-point function and the effective potential is maintained exactly
  also in a finite Taylor expansion for the numerical
  implementation. The flow equation for the sigma correlator involves up
  to four derivatives and thus requires $K\geq 4$.  We will employ $K=5$
  and $L=4$ in the following.

  \begin{table}[!b]
   \begin{center}
    \begin{tabular}{|c|c|c||c|c|c|}
     \hline
     $a/{\Lambda^2}$ & $b$ & ${c}/{\Lambda^3}$&$f_{\pi}$ &$m^\mathrm{scr}_{\pi}$&$m^\mathrm{scr}_{\sigma}$\\
     \hline
     -0.30&3.65&0.014&93.0&137.2&425.0\\
     \hline
     -0.34&3.40&0.002&93.1&16.4&299.8\\
     \hline
    \end{tabular}
   \end{center}
   \vspace{-.4cm}
   \caption{Parameter sets for a UV cutoff $\Lambda=500$~MeV corresponding to two different pion masses. The physical parameters, $f_{\pi}$ and the meson masses, are
   given in MeV.}  \label{tab:parameter}
  \end{table}
  
  The numerical results presented in this article were obtained using
  the Taylor-expansion method as described above, which by construction
  does not take into account the full field\hyph dependence of the effective
  potential. Hence we checked our results against the so-called
  grid method \cite{Bohr:2000gp}, which involves discretizing 
  the effective potential on a set of grid points in field space and
  solving the corresponding flow equations at fixed values in field
  space, from which the full field-depen\-dent effective potential can be
  reconstructed. The scale-dependent derivatives of the effective
  potential, evaluated at its minimum, can then be used as input for the
  corresponding flow equations (\ref{eq:flow2ptpisi}) for the 2-point
  functions. The disadvantage of the grid method is that one incurs
  considerably larger numerical costs for reaching low IR cutoff
  values than with the Taylor expansions. Fortunately, we were able to
  verify that the results from the grid method generally reproduce the
  Taylor results for $k \ge 0.05\Lambda$ very well. 
  
  The parameter sets used here for the $O(4)$ model with UV potential of
  the form $U_{k=\Lambda}=a\phi^2+b\phi^4$ are listed in
  Tab.~\ref{tab:parameter}.
    
  \subsection{Pion and sigma meson spectral functions}
\begin{figure}[!ht]
   \vspace*{.2cm}
   \centering \includegraphics[width=0.97\columnwidth]{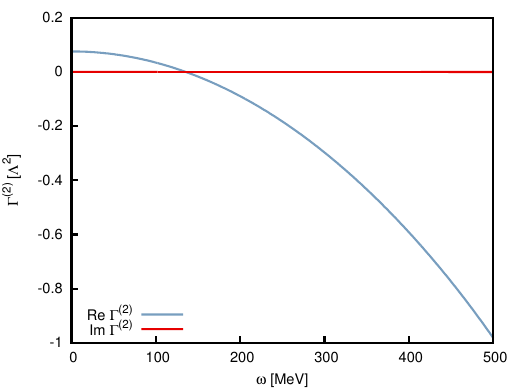}
   \vspace{-.2cm}
   \caption{Real and imaginary parts of the retarded pion propgator as a function
   of the time-like external momentum $\omega$.}  \label{fig:reimpi}
  \end{figure}

\begin{figure}[!ht]
   \vspace*{.2cm}
   \centering
   \includegraphics[width=0.97\columnwidth]{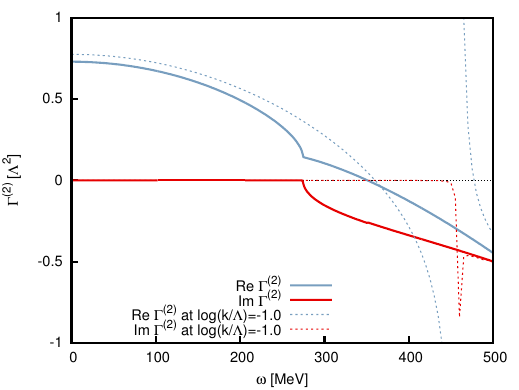}
   \vspace{-.2cm}
   \caption{Real and imaginary parts of the retarded sigma propagator as a function of
    the time-like external momentum $\omega$ at the IR scale and at an intermediate scale.}  \label{fig:reimsi}
  \end{figure}

  In \Fig{fig:reimpi} and \Fig{fig:reimsi} we show the real and
  imaginary parts of the retarded 2-point functions $\Gamma^{R}_\pi$ and
  $\Gamma^{R}_\sigma$, respectively. These were obtained by solving the
  flow equations (\ref{eq:flow2ptpisi}), separated into real and
  imaginary parts for a small but finite imaginary part $\epsilon \sim
  0.1$~MeV in the external frequency $\omega+i\epsilon$ as a function
  its real part $\omega$.
  For the pion the imaginary part in \Fig{fig:reimpi} remains
  negligible, whereas the real part behaves as demonstrated in earlier
  investigations \cite{Kamikado:2012bt,Strodthoff:2011tz}. The zero of
  the real part, which corresponds to the pion pole mass, is found at
  135.1~MeV as compared to a screening mass of 137.2~MeV. This once
  more reflects the difference between screening and pole masses which is a
  small effect of only a few percent, however, in the purely bosonic model
  \cite{Kamikado:2012bt,Strodthoff:2011tz}. At finite density,
  these values should be compared to the pion mass defined via the onset of pion
  condensation when coupling the model to an isospin chemical potential
   \cite{Svanes:2010we,Kamikado:2012bt}. The pole mass from our
   present truncation for the 2-point function is considerably closer to 
   the exact mass from the Bose-Einstein condensation transition at
   zero temperature, in particular, when fermions are included 
   as first observed in the Quark-Meson-Diquark model for two-color QCD 
   \cite{Strodthoff:2011tz}.  

  For the sigma meson, shown in \Fig{fig:reimsi}, the imaginary part
  remains negligible but only up to the 2-pion threshold at
  $\omega\approx2 m_\pi\approx 275$~MeV and tends to finite negative
  values thereafter. The two-pion emission threshold also shows up as
  a kink in the real part. The zero of the real part is found at
  350.6~MeV and we will argue that this value might be a good estimate
  for the sigma mass even in cases where no clear maximum in the
  corresponding spectral function is visible. For comparison, the
  value of the sigma screening mass extracted from the curvature of
  the effective potential results to be 425.0~MeV for the
  parameter set with nearly physical pion mass in
  Tab.~\ref{tab:parameter}.

  \begin{figure}[!b]
   \vspace*{.2cm}
   \includegraphics[width=0.97\columnwidth]{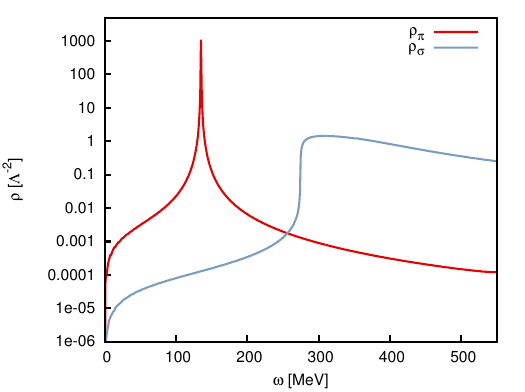}
   \vspace{-.2cm}
   \caption{Spectral functions for the pion ($\rho_\pi$) and the sigma meson ($\rho_\sigma$).}
   \label{fig:spectral}
  \end{figure}

  \Fig{fig:spectral} displays the pion and sigma spectral
  functions as calculated from \Eq{eq:spectralfn}. The pion spectral
  function shows a sharp peak at 135~MeV to be identified with the pion
  mass. The sigma spectral function exhibits the 2-pion threshold as a
  characteristic increase in the spectral density followed by a broad maximum
  at about 312~MeV above this threshold. Considering a case in which
  the imaginary part $\text{Im}\,\Gamma^{(2)}$ were independent of
  the (time-like) external momentum so that the only
  momentum-dependence was in the real part
  $\text{Re}\,\Gamma^{(2)}$, it is obvious from \Eq{eq:spectralfn} that
  the maximum of the spectral function would then occur at the zero of the real
  part with a width as determined by the constant imaginary
  part. This is the case for the pion 2-point function
  where the peak in the spectral function and zero of the 2-point
  function coincide as they must since the imaginary part practically
  vanishes. The same coincidence between the sigma mass defined via
  the maximum of the spectral function and the zero of the
  corresponding real part is of course no longer valid for a momentum-dependent
  imaginary part as obtained here, although the zero of the real part
  may still provide an estimate for the sigma mass. For the current
  parameter set the zero of the real part, however, overestimates the
  sigma mass defined via the maximum of the spectral function by 14\%. This 
  illustrates the difficulties of a mass assignment for a broad 
  resonance with a strongly momentum-dependent width. 
  In this case the imaginary part does not vanish at the zero of the real part. 
  Hence no zero is found in the complex plane of the physical sheet,
  consistent with the K\"allen-Lehmann representation, which
  allows only poles on the real axis. One expects a zero on the
  unphysical sheet, however, as illustrated in
  \cite{Hidaka:2002xv,Patkos:2002vr}.

  Except for a significantly lower sigma mass in our calculations, the
  zero-temperature spectral functions are in qualitative agreement with
  perturbation theory results
  \cite{Hidaka:2002xv,Chiku:2000eu,Chiku:1998kd,Chiku:1997va}. All
  these calculations show a sharp peak in the pion spectral function which
  determines the pion mass, and a resonance peak beyond the two-pion
  threshold in the sigma spectral function. This agreement is 
  reassuring since perturbation theory around the mean-field vacuum was
  found to work well outside the critical region. 
  For comparison, we
  have also performed corresponding perturbative calculations by
  integrating the flow equations for the 2-point functions with
  constant values for renormalized interactions and masses. While the
  qualitative features discussed above are indeed all present already
  in this simple one-loop approximation, the quantitative differences
  are quite considerable, however.  In particular, we generally 
  observe that the difference between pole and screening mass in the
  pion propagator, which is on the few percent level in the fully
  non-perturbative FRG solution as described above, increases
  considerably. With deviations on the order of around 100\%
  between the two, with pion pole masses typically almost twice as
  large as the input screening masses, the one-loop calculations
  therefore yield quantitatively rather inconsistent results for all
  input parameters we have considered.

  \begin{figure}[!ht]
   \vspace*{.2cm}
   \includegraphics[width=0.97\columnwidth]{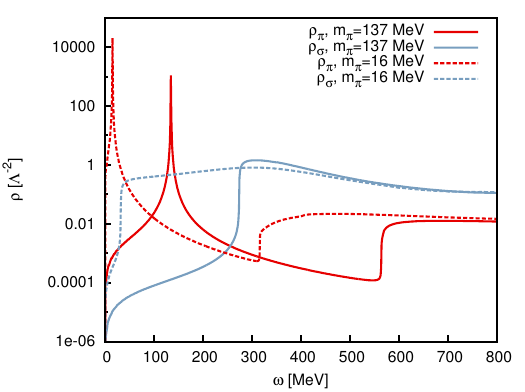}
   \vspace{-.2cm}
   \caption{Comparison of pion ($\rho_\pi$) and sigma meson ($\rho_\sigma$) spectral functions for parameter sets corresponding to two different pion masses.}
   \label{fig:spectralcomparisonmpi}
  \end{figure}

  As a consistency check we have also calculated the pion and
  sigma spectral functions closer to the chiral limit using the second
  parameter set in Tab.~\ref{tab:parameter}   
  corresponding to a pion pole mass of 16.0~MeV. The comparison with the
  physical pion mass is shown Fig.~\ref{fig:spectralcomparisonmpi}.
  When approaching the chiral limit, one expects the spectral
  weight of the pion pole in its correlator to increase more and more as
  this pole moves closer to the $\omega=0$ axis where it eventually 
  accumulates the full spectral weight in the chiral limit. Consequently, the
  spectral sum rule then implies that all other contributions to the
  spectral function should decrease with decreasing pion mass. Both
  these trends are seen in Fig.~\ref{fig:spectralcomparisonmpi} where
  we extended the frequency range of Fig.~\ref{fig:spectral} to
  include the $\sigma-\pi$ threshold in the pion spectral function.
  To explicitly verify the non-renormalization of the pion field in
  the chiral limit requires a more careful analysis, however,
  probably best done with polar coordinates in field space to correctly
  disentangle the Goldstone bosons from the radial mode \cite{Janprivatecomm}.
  
  \subsection{Spectral functions at intermediate scales}
  As described in the previous section, we find a resonance sigma at $k
  = 0$. It is illustrative to also consider the spectral functions at
  intermediate scales and thus their evolution with the RG scale $k$.
  \begin{figure}[!ht]
   \vspace*{.2cm}
   \includegraphics[width=0.97\columnwidth]{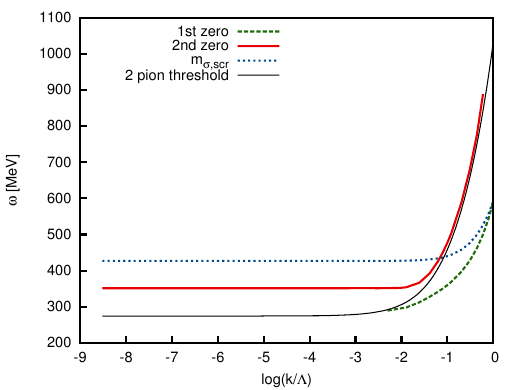}
   \vspace{-.2cm}
   \caption{Zeros of the real part of the sigma 2-point function, the
   sigma screening mass ($\sqrt{2U' + 4 \phi^2 U''}$) and the 2-pion
   threshold ($2\sqrt{k^2+2U'}$) as a function of the RG scale $k$.}
   \label{fig:zeros_intermediate}
  \end{figure}
  In Fig.~\ref{fig:zeros_intermediate}, we show the scale dependence of the
  zeros of the real part of the sigma's 2-point function together with
  its screening mass, defined via $\sqrt{2U' + 4\phi^2 U''}$, as well as 
  the 2-pion threshold at $2\sqrt{k^2+2U'}$.
  
  During the $k$-evolution two distinct zeros occur at
  intermediate scales. This is illustrated in Fig.~\ref{fig:reimsi}
  where we have also included the real and imaginary parts of
  $\Gamma^{(2)}_{\sigma\sigma}$ at such a scale (dashed lines). The
  first zero in the real part (dashed-blue line) represents the 
  pole corresponding to a renormalized sigma field. At the ultraviolet
  scale, the correlator $\Gamma^{(2)}_\sigma$ agrees with the classical
  one and its unique zero thus coincides with the screening mass of
  the bare sigma field. This zero gets renormalized by
  quantum fluctuations and starts to deviate from the sigma's screening
  mass at intermediate scales. It completely disappears at
  $\log(k/\Lambda) \sim -2.2$ when it reaches the 2-pion threshold.
  It turns out, however, that before this happens, a second zero emerges at
  larger $\omega $ (solid red line in Fig.~\ref{fig:zeros_intermediate}).
  This zero is dynamically generated via the
  $\sigma \rightarrow \pi + \pi$ process, and it is therefore not
  present at the ultraviolet cutoff scale yet. It arises above the
  scale dependent $\sigma \rightarrow \pi + \pi$ threshold at $\omega \sim
  2 E_{\pi}$. After decreasing alongside this threshold with the RG
  scale for a while, it approaches 
  its constant infrared value before the 2-pion threshold eventually
  does as well. Thus we find that, while the original
  zero disappears at some scale, a second one is generated and survives
  corresponding to the zero of the real part at the IR scale as shown in
  Fig.~\ref{fig:reimsi} (solid-blue line).

  \begin{figure}[!ht]
   \vspace*{.2cm}
   \includegraphics[width=0.97\columnwidth]{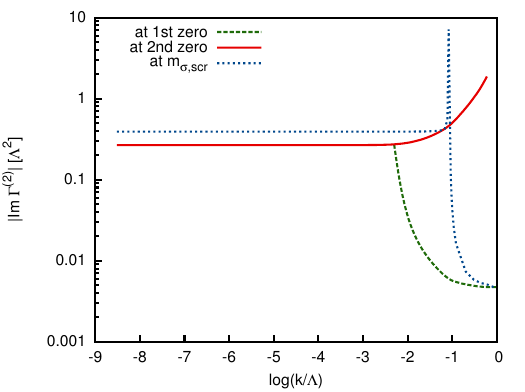}
   \vspace{-.2cm}
   \caption{Imaginary part of the retarded sigma propagator evaluated at the scale-dependent 1st zero
   (dashed green line), 2nd zero (solid red line) and sigma screening
   mass (dotted blue line).}  \label{fig:imaginary_sigma_intermediate}
  \end{figure}
  Fig.~\ref{fig:imaginary_sigma_intermediate} displays the imaginary
  part of the sigma correlator, evaluated at the first zero (dashed
  green line), second zero (solid red line) and the curvature mass
  (dotted blue line), respectively.  At large scales ($k \sim \Lambda$),
  both the first zero and the screening mass are below the
  scale-dependent 2-pion threshold and the corresponding imaginary parts
  vanish. The finite imaginary part in the numerical calculation is
  related to the finite imaginary external momentum employed to
  calculate the spectral function, see
  Eq.~(\ref{eq:analytic_continuation}).  Below $\log(k/\Lambda) \sim
  -1$, the screening mass crosses the threshold and the corresponding
  imaginary part of the propagator becomes finite. This behavior
  should be compared to the scale-dependence of the imaginary part in
  \cite{Floerchinger:2011sc}, which corresponds in some sense to an
  expansion of the 2-point function around the screening mass, albeit
  in a more refined truncation scheme. The second zero, which
  corresponds to the zero in the real part that survives until $k=0$,
  arises above the threshold and hence the imaginary part assumes a
  finite value. It decreases with $k$ and becomes constant below
  $\log(k/\Lambda) \sim -2$.

  These observations at intermediate scales indicate that the final zero in the real part of the sigma 2-point function does not emerge from the original
  single-particle contribution of the sigma at the cutoff scale by   
  renormalization effects,   
  but that it is predominantly due to the $\sigma \rightarrow \pi +
  \pi$ process. This is in contrast to the pion 2-point function whose
  single-particle contribution does flow continuously from the
  ultraviolet to the infrared and thus corresponds to the renormalized
  original pion mass. Such considerations question the
  reliability of truncation schemes based on an expansion around a 
  single scale-dependent pole \cite{Floerchinger:2011sc}. 
  It seems that the inclusion of the full external momentum dependence
  of the sigma 2-point function or at least an expansion around a more
  complicated singularity structure is required.

  \begin{figure}[!ht]
   \vspace*{.2cm}
   \includegraphics[width=0.97\columnwidth]{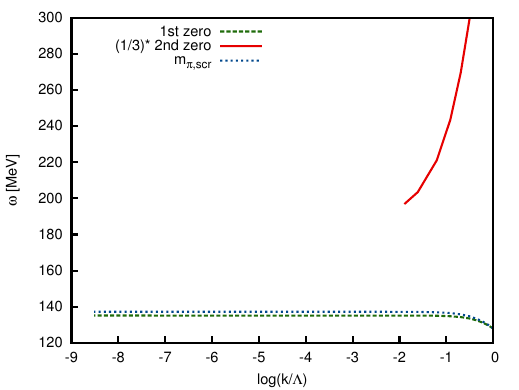}
   \vspace{-.2cm}
   \caption{Zeros of the real part of the pion 2-point function and the
   pion screening mass ($\sqrt{2U'}$) as a function of the RG scale
   $k$. The second zero was rescaled by a factor of $1/3$ to allow a comparison
   of all three quantities in a single figure.}  \label{fig:zeros_intermediate_pion}
  \end{figure}
  
  In order to contrast these conclusions for the sigma with
  the corresponding results for the pion correlator, we show in
  Fig.~\ref{fig:zeros_intermediate_pion} the analogue of 
  Fig.~\ref{fig:zeros_intermediate}, the zeros of the real part of the
  pion 2-point function. As before, we find two such zeros at
  intermediate scales, the first one (dashed green line) corresponding
  to the renormalized single-particle contribution from the pion
  screening mass at the UV scale, and the second (solid red line)
  corresponding to a dynamically generated pion 
  via the $\pi \rightarrow \pi + \sigma$ process. In this
  case the zero at $k=0$ is connected to the 
  first zero, however, whereas the second zero vanishes around $\log(k/\Lambda)
  \sim -1.9$. This underlines the fact that the zero in the pion 2-point
  function originates from the ultraviolet pion mass by finite renormalization.

 \section{Summary and Conclusions}
 In the present work we have extended the truncation scheme of
 Refs.~\cite{Strodthoff:2011tz,Kamikado:2012bt} for the 
 functional renormalization group (FRG) equations of 2-point
 functions. This scheme is consistent with the truncation for
 the effective potential and therefore ``symmetry conserving'' by construction. 
 In contrast to earlier approaches in
 the literature, we have employed an analytic continuation on the level of
 the flow equations which allows for a direct computation of retarded
 2-point functions at real frequencies. The numerical
 solutions of the corresponding flow equations yield realistic
 spectral functions of collective excitations, as we have demonstrated for
 the pion and the sigma meson in the $O(4)$ model. Their shape was
 found to be in good qualitative agreement with general expectations and
 previous studies. 

 To gain further insight into the origin of the pion and the sigma meson,
 we have evaluated the sigma and pion correlators at intermediate
 scales $k$. 
 It turns out that the pole in the pion propagator
 flows continuously from the initial screening mass at the ultraviolet cutoff
 to the physical single\hyph particle pole for $k\to 0$ by the
 renormalization effects. Since there is no renormalization of the
 pion in the chiral limit this ensures the existence of a zero-mode as
 required by Goldstone's theorem.  
 In stark contrast, the sigma pole at the infrared scale originates
 from a dynamical process. We have found two distinct trajectories in
 the singularity structure of the propagator at intermediate scales;
 one corresponding to the renormalized initial mass and the other to a
 dynamically generated complex pole on the unphysical sheet via the $\sigma 
 \rightarrow \pi + \pi$ coupling process. In particular, the sigma pole at $k=0$
 belongs to the branch which is dynamically generated. The other
 branch, connected to the bare sigma mass, disappears at some
 intermediate scale.
 
 We stress that it is possible to extend our approximation scheme
 to a thermal medium, and to include
 fermions. We are particularly interested in extending the framework to
 finite temperature and to investigate the medium-modified spectral functions.
 One should be able to observe the asymptotic restoration of chiral symmetry 
 also directly in the spectral functions. In the region around the
 phase boundary one 
 might expect to observe qualitative differences between the FRG and
 results based on perturbation theory.
     		       	 
 \section*{Acknowledgements}
 The authors thank Stefan Floerchinger, Jan Pawlowski and Arno Tripolt for useful
 discussions. K.~K.\ was supported by Grant-in-Aid for JSPS Fellows
 (No.~22-3671) and the Grant-in-Aid for the Global COE Program ``The Next
 Generation of Physics, Spun from Universality and Emergence.'' This work
 was supported by the Helmholtz International Center for FAIR within the
 LOEWE program of the State of Hesse and the European Commission,
 FP7-PEOPLE-2009-RG, No.~249203.

\appendix

\section{Breaking of Lorentz invariance}
  
In this Appendix we discuss the impact of the violation of Euclidean
O(4) resp.~Lorentz invariance by the employed three-dimensional
regulator function in Eq.~(\ref{eq:regulators}). On one hand such
regulator functions  
are particularly convenient as they do not introduce additional poles 
in the complex $p_0$-plane at finite RG scales $k$, and hence allow for a 
straightforward analytic continuation. On the other hand they have the 
slight disadvantage of breaking the underlying spacetime
symmetries. Perhaps more severely, however,  
they also allow arbitrarily large momentum transfers in the frequency 
direction which can potentially be problematic in combination with the
derivative expansion. Therefore, our present work should certainly 
be supplemented by computations using 4d-regulator functions in the
future. For now we can assess the systematic errors of our approach by
including finite spatial external momentum components in our flow
equations for the 2-pont functions. As an additional technical
complication, the spatial loop-momentum integrations can then no-longer
be perfomred analytically but have to be evaluated also numerically.

For Euclidean external 4-momenta, the O(4)-breaking effects induced by
the 3-dimensional regulator in the Euclidean 2-point functions are
truly negligible. We have compared the relative differences in their
integrated flows for spatial external momenta between zero and $|\vec
p|=500$~MeV and found that the Euclidean 2-point functions agree 
with each other within better than 1\% when plotted over the invariant
$\sqrt{p_0^2+\vec p^2}$ up to 300~MeV.

  \begin{figure}[!ht]
   \vspace*{.2cm}
   \includegraphics[width=0.97\columnwidth]{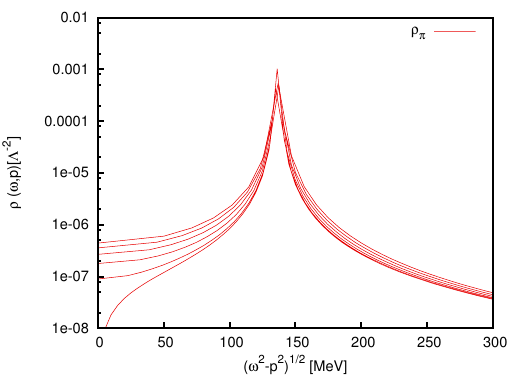}
   \vspace{-.2cm}
   \caption{Pion spectral function for Minkowski external momenta $\omega$ 
   and finite spatial momenta $p=0,50,\ldots,250$ MeV as function of 
   $\sqrt{\omega^2-\vec p^2}$ ($\epsilon=1$ MeV).}  \label{fig:checkmink}
  \end{figure}

For timelike external momenta the effect is somewhat more pronounced. 
In Fig.~\ref{fig:checkmink} we compare the spectral function of the
pion as obtained for different fixed values of the spatial external momentum
but now plotted as a function of the invariant $\sqrt{\omega^2-\vec
  p^2}$. Deviations of similar size are observed for the sigma
spectral function. Except in the very small momentum region where these
deviations are enhanced by the finiteness of our small residual imaginary part
$\epsilon$ in the frequency components (for the retarded boundary
conditions) the results are nevertheless in reasonable agreement with
the required Lorentz symmetry at vanishing temperature. 
The general shape and, most importantly the peak
position at the pion mass remain unchanged and are hence rather
robust results of our approach.
  
 \bibliographystyle{spphys.bst} 
 \bibliography{spectral}
\end{document}